\documentclass[fleqn,usenatbib,useAMS]{mnras}

\usepackage{adjustbox}
\usepackage{graphicx}	
\usepackage{amsmath}	
\usepackage{amssymb}	
\usepackage{multicol}        
\usepackage{bm}		
\usepackage{pdflscape}	

\usepackage[para]{threeparttable} 
\usepackage{url}
\usepackage{dcolumn}
\usepackage[para]{threeparttable} 
\usepackage[T1]{fontenc}
\usepackage{ae,aecompl}

\usepackage{newtxtext,newtxmath}

\title[Chemical analysis of NGC 6553]{Chemical analysis of the Bulge Globular Cluster NGC 6553}

\author[C. Montecinos et al.]{Carolina Montecinos$^{1}$
\thanks{E-mail: \href{mailto:caromontecinos@udec.cl}{caromontecinos@udec.cl}}
, S. Villanova$^{1}$, C. Mu\~{n}oz$^{1,3,4}$, C.C. Cort\'es$^{1,2}$.
\\
$^{1}$Departamento de Astronom\'ia, Casilla 160-C, Universidad de Concepci\'on, Concepci\'on, Chile.\\
$^{2}$Departamento de F\'isica, Facultad de Ciencias, Universidad del B\'io-B\'io, Avenida Collao 1202, Casilla 15-C, Concepci\'on, Chile.\\
$^{3}$Departamento de Astronom\'ia, Facultad de Ciencias, Universidad de La Serena. Avenida Juan Cisternas 1200, La Serena, Chile.\\
$^{4}$Instituto de Investigaci\'on Multidisciplinario en Ciencia y Tecnolog\'ia, Universidad de La Serena. Avenida Ra\'ul Bitr\'an S/N, La Serena, Chile.}

\date{Accepted XXX. Received YYY; in original form ZZZ}

\pubyear{2021}

\begin{document}
\label{firstpage}
\pagerange{\pageref{firstpage}--\pageref{lastpage}}
\maketitle

\begin{abstract}
Globular Clusters are among the oldest objects in the Galaxy, thus their researchers are key to understanding the processes of evolution and formation that the galaxy has experienced in early stages. Spectroscopic studies allow us to carry out detailed analyzes on the chemical composition of Globular Clusters. The aim of our research is to perform a detailed analysis of chemical abundances to a sample of stars of the Bulge Globular Cluster NGC 6553, in order to determine chemical patterns that allow us to appreciate the phenomenon of Multiple Population in one of the most metal-rich Globular Clusters in the Galaxy. This analysis is being carried out with data obtained by FLAMES/GIRAFFE spectrograph, VVV Survey and DR2 of Gaia Mission. We analyzed 20 Red Horizontal Branch Stars, being the first extensive spectroscopic abundance analysis for this cluster and measured 8 chemical elements (O, Na, Mg, Si, Ca, Ti, Cr and Ni), deriving a mean iron content of $[Fe/H] = -0.10\pm0.01$ and a mean of $[\alpha/Fe] = 0.21\pm0.02$, considering Mg, Si, Ca and Ti (errors on the mean). We found a significant spread in the content of Na but a small or negligible in O. We did not find an intrinsic variation in the content of $\alpha$ and iron-peak elements, showing a good agreement with the trend of the Bulge field stars, suggesting a similar origin and evolution.

\end{abstract}





\section{Introduction}

Theories about the formation of galaxies and how they evolve are still under study in the astrophysics (\citealp{2017MNRAS.472.2356M, 2018MNRAS.478.3994C}). The studies of our Galaxy and its three major components, i.e, the Bulge, the Disk and the Halo allow us to understand with more detail its formation and the evolution that it has experienced during its lifetime. For this reason, numerous surveys have been carried out with the purpose of improving our knowledge about the Milky Way. Some of them are: the Sloan Digital Sky Surveys (SDSS-IV;~\citealp{2019ApJS..240...23A}), VISTA Variables in the V\'ia L\'actea survey (VVV) \citep{2010NewA...15..433M}, RAdial Velocity Experiment survey (RAVE) \citep{2012sf2a.conf..121S}, the Gaia Mission \citep{2018AA...616A...4E} and in a near future the Large Synoptic Survey Telescope (LSST) \citep{2019BAAS...51g.268J}.

Within this context, the Bulge that is one of the oldest components of our galaxy with an age of $\thicksim10$ Gyr (\citealp{1995Natur.377..701O, 2002AJ....124.2054K, 2003AA...399..931Z, 2008ApJ...684.1110C, 2010ApJ...725L..19B, 2013AA...559A..98V, 2015ASPC..491..182G}), a metallicity range of $-1.5\lesssim[Fe/H]\lesssim+0.5$ and a formation timescale of $\thicksim2$ Gyr \citep{2018ARAA..56..223B}, is a fundamental region that helps us to reveal important information about the formation and evolution of our galaxy. Thus, despite the observational difficulties due to dust and reddening present in the center of the Milky Way, more and more researches are being carried out.

The globular clusters (GCs) located in the Bulge are representative of the ancient Bulge population and give information about its early formation. GCs are also interesting because for a long time they were thought to host only a single stellar population but over the last decade studies have revealed two (or more) stellar generations in a large number of galactic GCs, known today as multiple populations (MPs; \citealp{2009AA...505..117C, 2012AARv..20...50G}). The MPs are observed as a star-to-star variation in chemical abundances, a variation that is intrinsic of the stars and not due to stellar evolution \citep{2018ARAA..56...83B}. In GCs, stars present inhomogeneities in the content of their light elements involved in the proton-capture process, such as C, N, O, Na, Mg and Al (\citealp{1978ApJ...223..487C, 1992ApJ...395L..95D, 2001AJ....122.1438I, 2001AA...369...87G, 2004ARAA..42..385G, 2009AA...505..117C, 2009AA...505..139C, 2015AJ....149..153M, 2017MNRAS.472.2856S, 2019AJ....158...14N}). 

According to \citet{1989ATsir1538...11D, 1990SvAL...16..275D} all the elements that present a considerable abundance variation within GCs (C, N, O, Na, Mg and Al), originate during the Hydrogen burning at high temperatures, caused by the CNO, NeNa and MgAl cycles \citep{1993PASP..105..301L}.

To explain why these variations occur, the hypothesis of self-enrichment in GCs was proposed, which says that the presence of these chemical patterns is due to the self-enrichment that GCs suffered in their early stages of formation.

According to this hypothesis, the variation in light elements is due to the fact that a fraction of the observed stars were formed from material contaminated by ejecta of evolved stars (\citealp{2001AA...369...87G, 2004ARAA..42..385G, 2002AJ....123.3277R, 2006AA...458..135P, 2010AA...516A..55C, 2012ApJ...760...39P}). The nature of the polluting stars is still in debate but there are some candidates such as: massive asymptotic giant branch (AGB) stars (\citealp{2001ApJ...550L..65V, 2016ApJ...831L..17V, 2002AA...395...69D, 2016MNRAS.458.2122D, 2018MNRAS.475.3098D}), fast rotating massive main-sequence (MS) stars (\citealp{2006AA...458..135P, 2007AA...464.1029D, 2013AA...552A.121K}), massive MS binary stars (\citealp{2009AA...507L...1D, 2013MmSAI..84..171I}) and super massive stars \citep{2014MNRAS.437L..21D}.

The most famous signature of the presence of MPs in GCs are the anticorrelations between Sodium-Oxygen and Magnesium-Aluminum. These anticorrelations, especially the Na-O anticorrelation, have been presented in almost all the Galactic GCs studied to date (\citealp{2008AA...490..625M, 2009AA...505..117C, 2009AA...505..139C, 2010IAUS..266..326V, 2016MNRAS.460.2351V, 2017MNRAS.464.2730V, 2019ApJ...882..174V, 2012AARv..20...50G, 2013MNRAS.433.2006M, 2017MNRAS.466.1010S, 2018MNRAS.474.4541M, 2019MNRAS.483.1674R}). However, apparently there is a threshold both in mass and age for the appearance of the Na-O anticorrelation, since it has been observed mostly in old massive clusters (\citealp{2010AA...516A..55C, 2012AA...548A.122B}), with Ruprecht 106 being the only proved exception found until now \citep{2013ApJ...778..186V}. The less massive GC in which a Na-O anticorrelation was observed is NGC 6535 \citep{2017AA...607A..44B}, with a mass of $2.00\pm0.56 \times 10^{4}$ $M_{\odot}$ and an absolute visual magnitude of $M_{V} = -4.75$ (\citealp{2018MNRAS.478.1520B, 2010arXiv1012.3224H}). In addition, a variation in iron content has also been observed in some massive GCs, like Omega Centauri (\citealp{2014ApJ...791..107V, 2017MNRAS.469..800M}).

The Bulge GCs have not been studied in detail, mainly due to the high reddening and the field contamination. Thus, studies about MPs in Bulge GCs are not as extensive as in the Halo GCs. In this paper, we present a chemical analysis for NGC 6553, which is a metal-rich Bulge GC with a metallicity of $[Fe/H] = -0.18$, located in the constellation Sagittarius at 2.2 Kpc from the Galactic Center ($l = 5.25^\circ$, $b = -3.02^\circ$) and at 6.0 Kpc from the Sun \citep{2010arXiv1012.3224H}. This cluster have an age of $\thicksim 13$ Gyr \citep{2002ASSL..274..107Z}, a mass of $2.35\pm0.19 \times 10^{5}$ $M_{\odot}$ \citep{2018MNRAS.478.1520B}, an excess color of $E(B-V) =  0.63$ and an absolute visual magnitude of $M_{V} = -7.77$ \citep{2010arXiv1012.3224H}.

There have been several studies about NGC 6553. \citet{1999AA...341..539B} analyzed the abundance of two stars using The Cassegrain \'Echelle Spectrograph and found an overabundance of $\alpha$-elements, interpreting it as an evidence of a fast chemical enrichment, product of SNe Type II. \citet{1999ApJ...523..739C} studied abundance at 5 red horizontal branch (RHB) stars using the HIRES spectrograph of the Keck Telescope and found an excess of $\alpha$-elements together with a metallicity that reaches the average of the Galactic Bulge. \citet{2003AA...411..417M} perform the study of 5 giant stars using high resolution infrared spectra in the H band, obtained from Gemini-South 8m telescope, finding an enhancement of C+N and O concluding as well as \citet{1999AA...341..539B}
\begin{figure}
\begin{center}
 \includegraphics[width=0.5\textwidth]{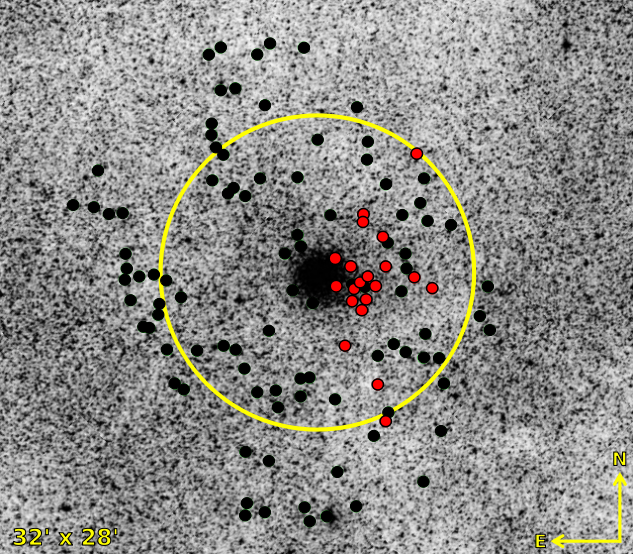}
 \caption{Spatial distribution of the observed stars in NGC 6553. Member stars are represented by red circles, while rejected stars are represented by black circles. The yellow line is the tidal radius of $\sim 8$ arcmin \citep{2010arXiv1012.3224H}.} 
 \label{fig:1}
 \end{center}
\end{figure}
that NGC 6553 was enriched by Type II SNe. Similar results were found for \citet{2006AA...460..269A}, who used high-resolution spectra of UVES spectrograph to perform an abundance analysis on 4 giant stars and \citet{2014AJ....148...67J} who used high-resolution spectra of FLAMES/GIRAFFE spectrograph to study the behavior of light, alpha and Fe-peak elements abundances at 12 stars. Studies more recently have been looking for signs of MPs in NGC 6553. \citet{2017MNRAS.466.1010S} used APOGEE data to perform a study of abundances at 12 stars, finding an extended anticorrelation between C and N. \citet{2017MNRAS.465...19T} used APOGEE data to study 10 stars, finding two star populations in C and N and \citet{2020MNRAS.492.3742M} studied 7 stars using the VLT/UVES spectrograph, finding a vertical Na-O anticorrelation, with a significant variation in Na, but not in O.

This paper will be organized as following: In section 2 we describe the observations and data reduction. In section 3 we explain the method used to determine stellar parameters, chemical abundances, and errors. In section 4 we present our results for alpha-elements, the Na-O anticorrelation and iron-peak elements. In section 5 we present the analysis of the orbit and finally in section 6 we present our conclusions.

\section{Observations and Data Reduction}

Our dataset consist of high-resolution spectra obtained with FLAMES/GIRAFFE spectrograph, mounted on UT2 Telescope (Kueyen) of ESO-VLT Observatory in Cerro Paranal (Chile), observed on July 2014 (ESO program ID 093.D-0286, PI S. Villanova). We used the set-up HR13, that gives a spectral coverage between 6112 {\AA} to 6400 {\AA}, with a central wavelength at $\sim 6273$ {\AA} and with a spectral resolution of $R \sim 22500$. We observed 116 stars, which were classified 69 like red giant branch (RGB) stars and 47 as horizontal branch (HB) sequence. The spatial distribution is shown in Figure \ref{fig:1}.  
\begin{figure}
 \includegraphics[width=0.5\textwidth, height=0.23\textheight]{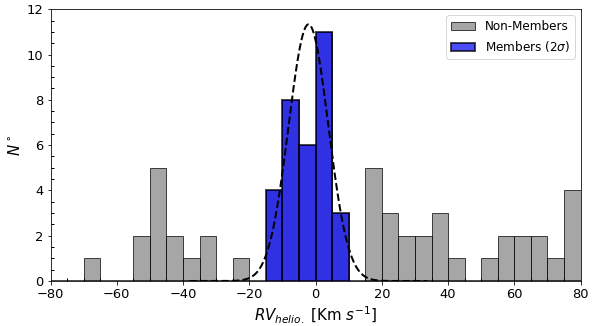}
 \caption{Histogram of heliocentric RVs for our observed stars. The blue histogram are the stars which were identified as members of the cluster according to RV only, where the dashed line is the Gaussian fit with a peak at $RV_{H}  = -2.13$ $Km$ $s^{-1}$. The gray histogram are the stars that were rejected.} 
 \label{fig:2}
\end{figure}

On the other hand, we also obtained near-infrared photometry data from VISTA Variables in the V\'ia L\'actea (VVV) survey \citep{2010NewA...15..433M} with a range between $K_{s} = 10.78$ mag and $K_{s} = 12.78$ mag and data obtained from Data Release 2 (DR2) of the Gaia Mission \citep{2018AA...616A...4E}. Data reduction was performed using the ESO CL based FLAMES/GIRAFFE Pipeline version 2.13.2\footnote{\url{http://www.eso.org/sci/software/pipelines/}}. Data reduction includes bias subtraction, flat-field correction, wavelength calibration and sky subtraction. We subtracted the sky using the {\ttfamily SARITH} package and measured radial velocities (RVs) using the {\ttfamily FXCOR} package, both in {\ttfamily IRAF} \footnote{IRAF is distributed by the National Optical Astronomy Observatory, which is operated by the Association of Universities for Research in Astronomy, Inc., under cooperative agreement with the National Science Foundation}.

 For observed RVs we used a synthetic spectrum as a template and corrected to the heliocentric system. The target stars were identified as cluster members if their heliocentric RVs were within $2\sigma$ from a mean value of $\langle RV_{H} \rangle  = -2.13$ $Km$ $s^{-1}$. As a result, cluster member stars are between $-15$ and $15$ $Km$ $s^{-1}$ with a velocity dispersion of $\sigma = 5.98$ $Km$ $s^{-1}$, as we clearly see in Figure \ref{fig:2}. Using RV we obtained 32 stars as members of the cluster, 7 RGB stars and 25 RHB stars.
\begin{figure}
 \begin{center}
 \includegraphics[width=0.48\textwidth, height=0.368\textheight]{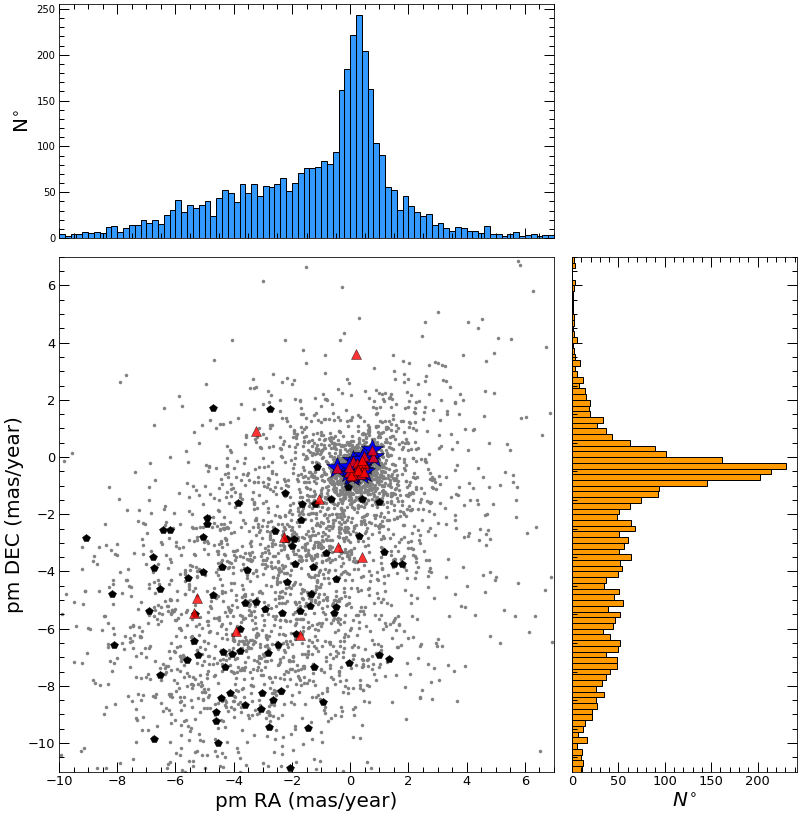}
 \caption{Proper motion diagram for the field stars and the globular cluster NGC 6553 stars, obtained from DR2 of Gaia. The histograms of 0.2  mas/yr bins in both directions, show the distribution of the stars with maximum points at $pm_{R.A.} = 0.19$ $mas/yr$ and $pm_{DEC} = -0.4$ $mas/yr$ respectively. In this diagram, the cluster is clearly distinguished from field stars. The gray dots are the field stars by Gaia. The black pentagons are rejected stars as members of the cluster. The red triangles are the members stars derived by RV only and the blue stars are the final clusters members found by proper motion.} 
 \label{fig:3}
 \end{center}
\end{figure}
\begin{figure}
 \begin{center}
 \includegraphics[width=0.46\textwidth]{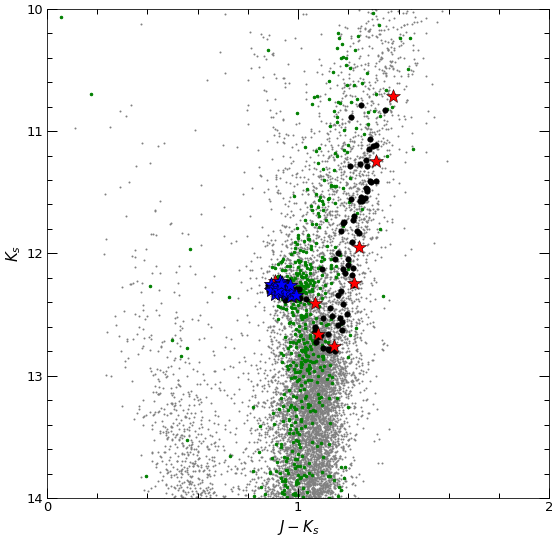}
 \caption{CMD of NGC 6553. The gray dots are the tidal radius ($\sim 8'$) of the cluster obtained from VVV photometry, corrected by the VVV reddening maps by \citet{2012AA...543A..13G}. The green dots are NGC 6553 stars (R < 1', this in order to distinguish its Red Clump). The observed stars are represented as black dots. The red stars are the members find by RV only and blue stars are the final members when also proper motions are considered.}
 \label{fig:4}
 \end{center}
\end{figure}
Then, in order to realise a more precise membership of the cluster, we combine the results obtained by RV with the proper motion of the stars obtained by the Gaia Mission\footnote{\url{https://gea.esac.esa.int/archive/}} \citep{2018AA...616A...4E}. For that, we perform a cross-match of our objects with the Gaia DR2 catalog. The result of this combination is reported in Figure \ref{fig:3}, where we plot the proper motion of NGC 6553. In this figure we can see that not all the stars identifies as members by RVs are actual members, but a fraction of them are just field stars because of their proper motions.

 Combining proper motions and RVs, we obtained 20 stars as members of the cluster with a mean proper motion of $pm_{R.A.}=0.28$ $mas$ $year^{-1}$ and $pm_{DEC}=-0.35$ $mas$ $year^{-1}$. The mean heliocentric RV value for the final member stars is $\langle RV_{H} \rangle  = -3.33\pm1.16$ $Km$ $s^{-1}$, while the dispersion is 5.44 $Km$ $s^{-1}$. This value is in agreement with $\langle RV_{H} \rangle = -3.2$ $Km$ $s^{-1}$ \citep{2010arXiv1012.3224H}, $\langle RV_{H} \rangle = 4 \pm 7.1$ $Km$ $s^{-1}$ \citep{1999ApJ...523..739C}, $\langle RV_{H} \rangle = 1.6 \pm 6$ $Km$ $s^{-1}$ \citep{2003AA...411..417M}, $\langle RV_{H} \rangle = -1.86 \pm 2.01$ $Km$ $s^{-1}$ \citep{2006AA...460..269A}, $\langle RV_{H} \rangle = -0.14 \pm 5.47$ $Km$ $s^{-1}$ \citep{2017MNRAS.465...19T} and $\langle RV_{H} \rangle = -3.86 \pm 2.12$ $Km$ $s^{-1}$ \citep{2020MNRAS.492.3742M}.
 \begin{table*}
 \caption{Coordinates, 2MASS (J, H, $K_{s}$) magnitudes and radial heliocentric velocities for the observed stars. The J, H and $K_{s}$ photometry are from VVV-PSF photometry  (\citealp{2014AA...563A..76M, 2017MNRAS.464.1874C}).}
 \label{tab:1}
 \resizebox{13cm}{!} {
 \begin{tabular}{llllllllllll}\\
  \hline
  Star & RA & DEC & $J$ & $H$ & $K_{s}$ & $RV_{H}$ &\\
    & [degrees] & [degress] & [mag] & [mag] & [mag] & [$Kms^{-1}$]&\\
  \hline
  01 & 272.259083 & -26.034027 & 13.2897 & 12.5248 & 12.3193 & +1.96&\\[2pt]
  
  02 & 272.266624 & -26.002666 & 13.1558 & 12.4248 & 12.2275 & -10.29&\\[2pt]
  
  03 & 272.215708 & -25.921111 & 13.2825 & 12.5204 & 12.3299 & -3.61&\\[2pt] 

  04 & 272.230458 & -25.807638 & 13.1429 & 12.4256 & 12.2504 &  +3.26&\\[2pt]

  05 & 272.231749 & -25.912277 & 13.2605 & 12.5344 & 12.3139 & +1.91&\\[2pt] 
  
  06 & 272.258958 & -25.902694 & 13.1892 & 12.4985 & 12.2926 &  -2.97&\\[2pt] 

  07 & 272.261541 & -25.87775  & 13.2535 & 12.5074 & 12.3199 &  -0.44&\\[2pt]

  08 & 272.267958 & -25.9195   & 13.2385 & 12.5044 & 12.3307 &  +2.50 &\\[2pt] 
  
  09 & 272.275666 & -25.912111 & 13.1805 & 12.4314 & 12.2447 & -5.19&\\[2pt]

  10 & 272.277833 & -25.931111 & 13.3225 & 12.5584 & 12.3497 &  +1.15&\\[2pt]

  11 & 272.280708 & -25.865527 & 13.1902 & 12.4575 & 12.2976 &  -5.56&\\[2pt] 

  12 & 272.280916 & -25.85875  & 13.1812 & 12.4545 & 12.2736 & -6.47&\\[2pt] 

  13 & 272.281374 & -25.939861 & 13.2575 & 12.4994 & 12.3057 & -13.25&\\[2pt] 

  14 & 272.282583 & -25.917388 & 13.2735 & 12.5104 & 12.3107 &  -12.74&\\[2pt] 

  15 & 272.289291 & -25.921166 & 13.2605 & 12.5354 & 12.3187 & -0.56&\\[2pt]

  16 & 272.291124 & -25.932527 & 13.2073 & 12.455  & 12.2826 &  -4.92&\\[2pt] 

  17 & 272.291958 & -25.903083 & 13.331  & 12.5601 & 12.3403 &  -10.07&\\[2pt]
  
  18 & 272.297749 & -25.970333 & 13.217  &12.457   & 12.2954 &  -8.54&\\[2pt] 

  19 & 272.305458 & -25.919888 & 13.2293 & 12.479  & 12.2606 &  -8.63&\\[2pt] 

  20 & 272.306791 & -25.896166 & 13.1792 & 12.4242 & 12.2476 &  +4.03&\\[2pt] 

  \hline
 \end{tabular}
 }
\end{table*}

The star which is outside of tidal radius in Figure \ref{fig:1} corresponds to the star $N^{\circ} 01$ of Tables \ref{tab:1} and \ref{tab:3}, where we can see that both its RV as well as its photometry and chemical abundance are in agreement with the rest of the sample, therefore, we have considered it as a member of the cluster.

Figure \ref{fig:4} show the color-magnitude diagram (CMD) of NGC 6553. In the CMD we can see that all the member stars belong to the HB sequence, while the RGB stars are Bulge field stars.

Table \ref{tab:1} list the basic parameters of the members stars, i.e., the ID, J2000 coordinates (RA and Dec in degrees), J, H, $K_{s}$ magnitudes from VVV-PSF photometry, calibrated on the system of 2MASS (\citealp{2014AA...563A..76M, 2017MNRAS.464.1874C}) and heliocentric radial velocity ($RV_{H}$).

The determination of the stellar parameters is discussed in the next section.

\section{Atmospheric parameters, Abundances and Errors}

\subsection{Atmospheric parameters}

The Globular Cluster NGC 6553 is located in the Galactic Bulge (GB), a region which houses the highest star density of the galaxy together with gas and dust that obscures the view, that making difficult its observation. In order to compare different methodologies, we decided to obtain atmospheric parameters both from the spectra and from the photometry.

\begin{table*}
 \caption{Spectroscopic and Photometric stellar parameters}
 \label{tab:2}
 \resizebox{14cm}{!} {
\begin{tabular}{lllllllllllr}
\firsthline
\hline
\multicolumn{6}{c}{Spectroscopic parameters} & \multicolumn{5}{c}{Photometric parameters}\\
\cline{2-6}
\cline{8-11}
Star & $T_{eff}$ & $log(g)$ & $\upsilon_{t}$ & [Fe/H] & FeI/FeII & & $T_{eff}$ & $log(g)$ & $\upsilon_{t}$ & [Fe/H]&\\
& [K] & [dex] & [$Kms^{-1}$] & [dex] & & & [K] & [dex] & [$Kms^{-1}$] & [dex]\\
\hline
  01 & 4870 & 2.20 & 1.75 & -0.06 & 26/3 & & 4780 & 2.11 & 1.68 & -0.10 &\\[2pt]
  
  02 & 4879 & 2.52 & 1.71 & -0.18 & 24/3 & & 4770 & 2.40 & 1.61 & -0.18 &\\[2pt]
  
  03 & 4752 & 2.30 & 1.46 & -0.08 & 24/4 & & 4691 & 2.22 & 1.43 & -0.11 &\\[2pt] 

  04 & 4650 & 2.00 & 1.54 & -0.06 & 21/2 & & 4600 & 2.00 & 1.53 & -0.08 &\\[2pt]

  05 & 4850 & 2.30 & 1.68 & -0.05 & 26/3 & & 4844 & 2.20 & 1.67 & -0.05 &\\[2pt] 
  
  06 & 4837 & 2.49 & 1.48 & -0.14 & 20/4 & & 4780 & 2.40 & 1.47 & -0.18 &\\[2pt] 

  07 & 4900 & 2.44 & 1.49 & -0.07 & 18/4 & & 4764 & 2.31 & 1.33 & -0.07 &\\[2pt]

  08 & 4396 & 1.62 & 1.11 & -0.16 & 20/3 & & 4680 & 2.18 & 1.33 & -0.07 &\\[2pt]
  
  09 & 4720 & 2.08 & 1.35 & -0.11 & 26/3 & & 4650 & 2.05 & 1.28 & -0.11 &\\[2pt]

  10 & 4980 & 2.50 & 1.87 & -0.14 & 26/3 & & 4875 & 2.39 & 1.73 & -0.15 &\\[2pt]

  11 & 4575 & 1.65 & 1.27 & -0.18 & 20/2 & & 4464 & 2.05 & 1.07 & -0.04 &\\[2pt]

  12 & 4744 & 2.23 & 1.24 & -0.04 & 23/3 & & 4665 & 2.08 & 1.24 & -0.12 &\\[2pt]

  13 & 4455 & 2.00 & 1.32 & -0.07 & 18/3 & & 4860 & 2.41 & 1.70 & -0.06 &\\[2pt] 

  14 & 4840 & 2.50 & 1.58 & -0.17 & 26/3 & & 4810 & 2.42 & 1.54 & -0.18 &\\[2pt] 

  15 & 4770 & 2.40 & 1.75 & -0.19 & 23/3 & & 4731 & 2.46 & 1.69 & -0.19 &\\[2pt]

  16 & 4720 & 2.30 & 1.60 & -0.08 & 24/3 & & 4760 & 2.15 & 1.61 & -0.06 &\\[2pt] 

  17 & 4752 & 2.40 & 1.26 & -0.04 & 21/3 & & 4645 & 2.08 & 1.16 & -0.06 &\\[2pt]
  
  18 & 4675 & 2.10 & 1.48 & -0.05 & 29/4 & & 4542 & 2.00 & 1.33 & -0.04 &\\[2pt] 

  19 & 4778 & 2.10 & 1.55 & -0.04 & 20/3 & & 4667 & 2.12 & 1.45 & -0.05 &\\[2pt] 

  20 & 4825 & 2.40 & 1.67 & -0.10 & 25/3 & & 4685 & 2.13 & 1.46 & -0.07 &\\[2pt]
\lasthline
\end{tabular}
}
 \begin{flushleft}
 \footnotetext \footnotesize {\small {{\textbf{NOTE.} Column 6 are the numbers of lines measured for FeI and FeII.}}} \\
 \end{flushleft}
\end{table*}

\subsubsection{Photometric stellar parameters}

The estimation for the photometric stellar parameters was obtained in the following way. $T_{eff}$ was derived using the color-temperature relation of \citet{1999AAS..140..261A}. For that, we translate the (J-K) color form the 2MASS system to CIT (California Institute of Technology), and from this to TCS (Telescopio Carlos S\'anchez), using the relations from \citet{1998AAS..131..209A}. On the other hand, due to the fact that this GC is in the GB, we have to consider the differential reddening, which was obtained using the Bulge Extinction And Metallicity calculator\footnote{\url{http://mill.astro.puc.cl/BEAM/calculator.php}} developed by \citet{2012AA...543A..13G}. Surface gravity, log(g), was obtained from the canonical equation:

\begin{equation}
\large{\log \left(\frac{g}{g_{\odot}}\right) = \log \left(\frac{M}{M_{\odot}}\right) + 4\log \left(\frac{T_{eff}}{T_{\odot}}\right) - \log \left(\frac{L}{L_{\odot}}\right)},    
\end{equation}
where the mass M was assumed to be 0.5 $M_{\odot}$, which is the typical mass of RHB stars in globular clusters and the luminosity $L/L_{\odot}$ was obtained from the absolute magnitude $M_{Ks}$ assuming an intrinsic distance module of $(m-M)_{0} = 13.88$, obtained from $(m-M)_{V} = 15.83$ and E(B-V) = 0.63 from \citet{2010arXiv1012.3224H}. The bolometric correction (BC) was derived by adopting the relation BC-$T_{eff}$ from \citet{{2010MNRAS.403.1592B}}. Finally, microturbulence velocity ($\upsilon_{t}$) was obtained from the relation of \citet{2008AA...490..625M}
\begin{equation}
    \upsilon_{t} = 2.22 - 0.322 \cdot log(g)
\end{equation}

Then, atmospheric models are calculated, adopting the values of $T_ {eff}$, log (g) and $\upsilon_ {t}$ obtained before and using [Fe/H] = -0.18 from \citet{2010arXiv1012.3224H}. Finally the metallicity of each star was obtained using MOOG \citep{1973ApJ...184..839S}, which is a program that realize analysis assuming a Local Thermodynamic Equilibrium (LTE) and the line list for FeI described in \citet{villanova}.

\subsubsection{Spectroscopic stellar parameters}

First we calculated an atmospheric model using ATLAS9 \citep{1970SAOSR.309.....K} taking initial guesses for $T_{eff}$, $log(g)$, [Fe/H] and $\upsilon_{t}$. Then we adjusted this values in the program iteratively in order to remove any trends in Excitation Potential (E.P.) and Equivalent width (EW) vs abundance of the Fe$_{I}$ lines to obtain the final $T_{eff}$ and $\upsilon_{t}$ respectively. The final $log(g)$ was obtained satisfying the ionization equilibrium of the Fe$_{I}$ and Fe$_{II}$ lines. Due to the limited spectral range of our GIRAFFE observations, we used 18/29 Fe$_{I}$ and 2/4 Fe$_{II}$ lines for the latter purpose (see Table \ref{tab:2}). Finally, the value of [Fe/H] was change in each iteration by the value of the output of the abundance analysis.

To calculate the atmospheric parameters, we using MOOG \citep{1973ApJ...184..839S}.The line list for the chemical analysis is the one described in \citet{villanova}.
\begin{figure}
 \includegraphics[width=0.52\textwidth, height=0.346\textheight]{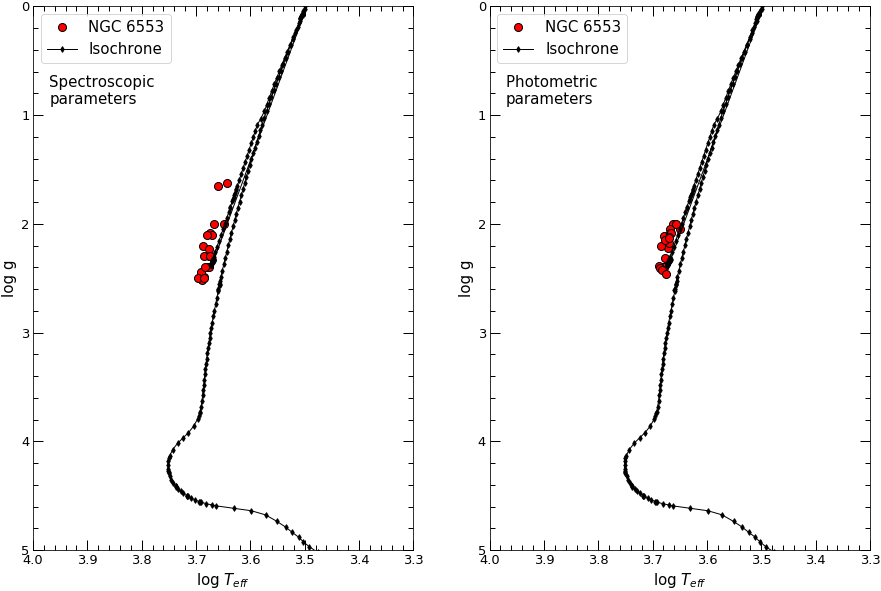}
 \caption{Log ($T_{eff}$) vs Log(g) for our sample of stars. The plotted isochrone has a metallicity of $[Fe/H] = - 0.15$ dex, $[\alpha/Fe] = +0.13$ dex and age of 13 Gyr \citep{2017MNRAS.465...19T}.} 
 \label{fig:5}
\end{figure}

Figure \ref{fig:5} shows a comparison between the spectroscopic and photometric parameters with an isochrone of 13 Gyr at a similar metallicity [Fe/H] = -0.15 dex \citep{2017MNRAS.465...19T}. A small offset is observed between our data and isochrone, of the order of $\thicksim$ 100 K. In any case the agreement between spectroscopic and photometric temperatures is good \citep{2013MNRAS.430..836N}. On the other hand, it is also observed that with the spectroscopic method, log (g) span a range of 1 dex, with a mean of 2.23 dex and a rms of 0.26 dex. On the other hand, with the photometric method, log(g) span a range of 0.46 dex, with a mean of 2.21 dex and a rms of 0.15 dex.

The final spectroscopic and photometric parameters are listed in Table \ref{tab:2}.

\subsection{Abundances} 

\begin{figure}
 \includegraphics[width=0.54\textwidth, height=0.58\textheight]{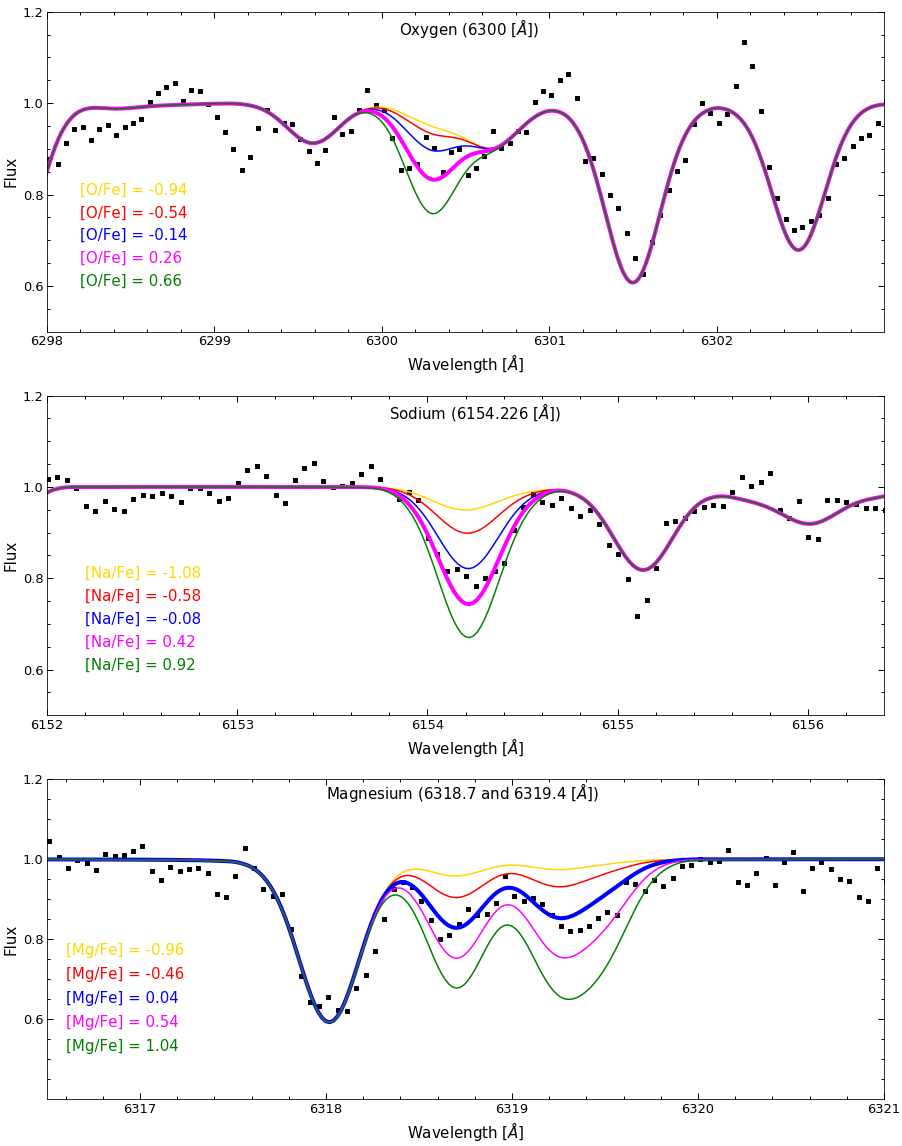}
 \caption{Oxygen, Sodium and Magnesium lines fitting process in star {\#}09. Dotted line correspond to the observed spectrum and solid color lines are synthesized spectra for different abundances. The thicker line correspond to the best fit model.}
 \label{fig:6}
\end{figure}

We use the spectroscopic parameters to derive the chemical abundances of the RHB stars. For the well-defined spectral lines  (Si, Ca, Ti, Cr and Ni), the abundances were obtained from equivalent width (EW's). The detailed explication of the method use to measure the EW's is given in \citet{2008AA...490..625M}. For the others elements, whose lines were affected by blending or were weak (O, Na and Mg), we used the spectrum-synthesis method. In the spectrum-synthesis method we calculated 5 synthetics spectra for different abundances and compares them with the observed spectra, where we consider all the lines in a range of $\pm5${\AA}. Then, we choose as the best fit, the abundance that minimises rms scatter. For this method, we use a line list that contains all the important spectral lines, where for each line to be analyzed, we selected a region of $\pm20$ {\AA} centered on the line itself. An example of this method is plotted in Figure \ref{fig:6} where the observed spectrum of the star \#09 is compared with synthetic spectra for O, Na and Mg. For the continuum determination, we selected a range of $\pm10$ {\AA} around the line, but in the Figure \ref{fig:6} we just show a range of  $\pm2$ {\AA}. O line was carefully treated since it was heavily affected by the corresponding telluric line in emission.

Oxygen abundances were obtained from the forbidden [OI] line at 6300 {\AA}. The forbidden [OI] line at 6363 {\AA} was as much or more affected by the corresponding telluric line in emission than the [OI] line at 6300 {\AA}, therefore we have not considered it for this work. Sodium abundances were obtained from  the NaI  doublet at 6154 - 6160 {\AA} and Magnesium abundances were derived from the MgI doublet at 6318.7 and 6319.4 {\AA}.

The iron abundances were obtained from EW's method using 18-29 lines for FeI and 2-4 for FeII. The adopted solar abundances are from \citet{2013ApJ...778..186V}. In the Table \ref{tab:3} we are report the chemical abundances that we measure and the adopted solar abundance values.\\

\begin{table*}
 \caption{Abundances of the observed stars for NGC 6553.}
 \label{tab:3}
 \resizebox{18cm}{!} {
 \begin{tabular}{lllllllllll}\\
  \hline
  \hline
   Star & [O/Fe] & $[Na/Fe]_{NLTE}$ & [Mg/Fe] & [Si/Fe] & [Ca/Fe] & [Ti/Fe] & [Cr/Fe] & [Fe/H] & [Ni/Fe] & [$\alpha$/Fe]\\
  \hline  
  \hline
  01 & -0.88 & +0.17 & +0.46 & -0.02 & +0.34 & +0.17 & -0.07 & -0.06 & +0.11 & +0.24 \\ \\[2pt] 
  
  02 & - & +0.52 & +0.52 & +0.19 & +0.33 & +0.71 & - & -0.15 & +0.29 & +0.44 \\ \\[2pt] 
  
  03 & -0.61 & +0.6 & +0.31 & +0.09 & +0.53 & +0.66 & +0.07 & -0.08 & +0.14 & +0.40\\ \\[2pt] 
  
  04 & -0.28 & +0.18 & +0.21 & +0.14 & +0.45 & +0.39 & -0.50 & -0.06 & +0.09 & +0.30 \\ \\[2pt] 

  05 & +0.14 & +0.19 & +0.26 & +0.19 & -0.03 & +0.31 &  0.00 & -0.04 & +0.01 & +0.18 \\ \\[2pt]
  
  06 & -0.64 & +0.61 & +0.10 & -0.10 & +0.32 & +0.28 & - & -0.14 & +0.12 & +0.15 \\ \\[2pt] 

  07 & -0.94 & +0.03 & +0.38 & -0.05 & +0.34 & +0.51 & +0.13 & -0.07 & +0.14 & +0.30 \\ \\[2pt]

  08 & -0.48 & -0.11 & -0.01 & +0.38 & -0.19 & -0.11 & -0.88 & -0.13 & +0.13 & +0.02 \\ \\[2pt]
  
  09 & +0.29 & +0.41 & +0.05 & +0.01 & +0.64 & +0.21 & 0.00 & -0.11 & +0.20 & +0.23 \\ \\[2pt]

  10 & +0.29 & +0.42 & +0.28 & 0.00 & +0.26 & +0.49 & +0.13 & -0.14 & +0.07 & +0.26 \\ \\[2pt]

  11 & - & +0.39 & +0.04 & +0.12 & +0.31 & +0.13 & -0.63 & -0.17 & +0.23 & +0.15 \\ \\[2pt] 

  12 & -0.94 & +0.45 & +0.02 & +0.19 & +0.33 & +0.51 & - & -0.04 & +0.18 & +0.26 \\ \\[2pt]

  13 & -0.11 & -0.10 & -0.01 & +0.36 & -0.09 & +0.12 & -0.23 & -0.07 & +0.13 & +0.01 \\ \\[2pt] 

  14 & +0.3 & +0.51 & +0.31 & +0.16 & +0.18 & +0.25 & -0.13 & -0.18 & +0.19 & +0.23 \\ \\[2pt] 

  15 & -0.93 & +0.63 & +0.33 & +0.09 & -0.05 & +0.27 & -0.06 & -0.20 & +0.20 & +0.16 \\ \\[2pt] 

  16 & +0.64 & +0.11 & +0.05 & -0.05 & +0.09 & +0.05 & -0.10 & -0.07 & -0.04 & +0.03 \\ \\[2pt]

  17 & - & +0.45 & +0.07 & +0.23 & +0.12 & +0.36 & -0.29 & -0.03 & +0.23 & +0.20 \\ \\[2pt]
  
  18 & - & +0.15 & +0.18 & +0.13 & +0.33 & +0.33 & -0.40 & -0.05 & +0.19 & +0.24 \\ \\[2pt] 

  19 & -0.29 & +0.27 & +0.09 & +0.12 & - & +0.38 & -0.22 & -0.03 & 0.00 & +0.20 \\ \\[2pt] 

  20 & -0.95 & +0.44 & +0.10 & +0.17 & +0.13 & +0.24 & -0.20 & -0.09 & +0.06 & +0.16 \\ \\[2pt] 
  
 \hline 
 \hline 
  Cluster & $-0.34\pm 0.13$ & $+0.31\pm 0.05$ & $+0.19\pm 0.03$ & $+0.12\pm 0.03$ & $+0.23\pm 0.05$ & $+0.31\pm 0.04$ & $-0.20\pm 0.06$ & $-0.10\pm 0.01$ & $+0.13\pm 0.02$ & $0.20\pm 0.02$ \\[2pt] 
  
  Sun & 8.83 & 6.32 & 7.56 & 7.61 & 6.39 & 4.94 & 5.63 & 7.50 & 6.26 & \\[2pt] 
\lasthline
 \end{tabular}
}
\end{table*}

\subsection{Errors}

\begin{figure}
 \includegraphics[width=0.53\textwidth, height=0.17\textheight]{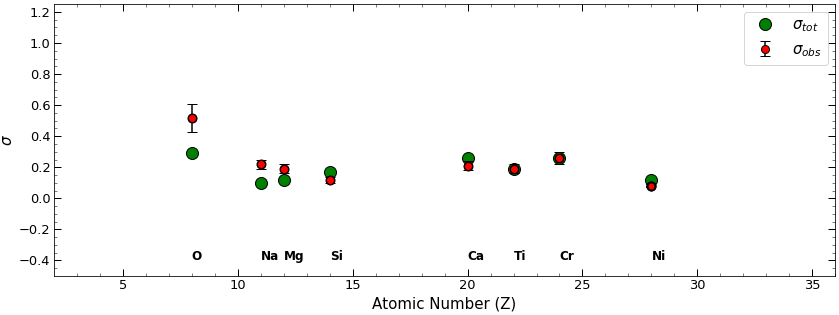}
 \caption{A comparison between $\sigma_{tot}$ and $\sigma_{obs}$ as function of the Atomic Number (Z). Green circles are the total errors of our measurements while the red circles are the observed errors. The error bars represent the error of the mean.} 
 \label{fig:7}
\end{figure}

The internal error analysis associated with our spectroscopic stellar parameters were performed in the following way. First we calculated the typical error on the slope in the relation abundance vs E.P., corresponding to the average of the errors associated for each star. Then, we choosing a star (\#14) as representative of the entire sample. Afterwards, to calculate an estimate of the error in temperature, we fixed the other parameters and varied the temperature until the slope of the line that best fits the relation abundance vs E.P. became equal to the respective mean error. For $\upsilon_{t}$, the same procedure was applied, but using the relation between abundance and EWs. Then, to calculate an estimate of the error in  $log(g)$, we varied the gravity until satisfied the relation:

{\begin{equation}
[FeI/H] - \overline{\sigma_{\star[FeI/H]}} = [FeII/H] + \overline{\sigma_{\star[FeII/H]}}   
\end{equation}}

where ${\sigma_{\star[Fe/H]}}$ is the dispersion of the Fe abundances divided by $\sqrt{N_{lines} - 1}$. Finally, in order to associate an error to [Fe/H], we divided the $\sigma_{obs}$, which is the observed dispersion, by $\sqrt{N_{lines} - 1}$. In this way, we obtained the following values, $\Delta T_{eff}$ = +50K, $\Delta log(g)$ = +0.2 dex, $\Delta [Fe/H]$ = +0.1 dex and $\Delta \upsilon_{t}$ = +0.1 [$Kms^{-1}$]. Then, we varied the $T_{eff}$, log(g), [Fe/H] and $\upsilon_{t}$ of the star \#14 according with the atmospheric errors that we obtained before and redetermine the abundances as in \citet{2008AA...490..625M}. Afterwards, we measured the error due to noise in the spectra ($\sigma_{S/N}$) for the elements measured by EW's and by spectrum-synthesis. The elements whose abundance was obtained by EW's, the error was obtained dividing the average rms scatter of the FeI lines given by MOOG by the square root of the number of lines used for a given element and a given star. On the other hand, for the elements whose abundance was obtained by spectrum-synthesis the error was obtained comparing the scatter of the observed spectrum with the synthetic best fitting spectrum. This gives the S/N error for each line. The S/N error for each element is obtained dividing this value for the root square of the number of lines used for the element. Finally, the total error ($\sigma_{Tot}$) in our measurements of abundances, is given by the relation:

{\begin{equation}
\sigma_{tot} = \sqrt{\sigma^2_{T_{eff}} + \sigma^2_{log(g)} + \sigma^2_{\upsilon_{t}} + \sigma^2_{[Fe/H]} + \sigma^2_{S/N}}  \end{equation}}

The error for each [X/Fe] ratio as result of uncertainties in the spectroscopic stellar parameters and $\sigma_{S/N}$ are reported in the Table \ref{tab:4}. In the Figure \ref{fig:7} we plot the total and observed errors as function of the atomic number for each element. The error bars represents the error of the observed dispersion ($rms/ \sqrt{2N}$, with N the number of stars).
\begin{table*}
\caption{Estimated errors on abundances, due to errors on atmospheric parameters and spectral noise, compared with the observed errors.}
\label{tab:4}
\resizebox{18cm}{!}{
\begin{tabular}{lllllllllll}\\
\hline
\hline
ID & $\Delta T_{eff} = 50$K & $\Delta log(g) = 0.2$ & $\Delta \upsilon_{t} = 0.1$ & $\Delta [Fe/H] = 0.1$ & $\sigma_{S/N}$ & $\sigma_{tot}$ & $\sigma_{obs}$ & \\
\hline  
\hline
$\Delta([O/Fe])$ & 0.04 & 0.07 & 0.04 & 0.06 & 0.27 & 0.29 & $0.52\pm0.09$ &\\[2pt] 
  
$\Delta([Na/Fe])$ & 0.00 & 0.01 & 0.03 & 0.00 & 0.10 & 0.10 & $0.22\pm0.03$ &\\[2pt] 
  
$\Delta([Mg/Fe])$ & 0.01 & 0.00 & 0.03 & 0.06 & 0.10 & 0.12 & $0.19\pm0.03$ & \\[2pt] 

$\Delta([Si/Fe])$ & 0.06 & 0.04 & 0.04 & 0.00 & 0.15 & 0.17 & $0.12\pm0.02$ & \\[2pt]

$\Delta([Ca/Fe])$ & 0.01 & 0.03 & 0.00 & 0.03 & 0.26 & 0.26 & $0.21\pm0.03$ & \\[2pt] 
  
$\Delta([Ti/Fe])$ & 0.03 & 0.01 & 0.00 & 0.03 & 0.18 & 0.19 & $0.19\pm0.03$ & \\[2pt] 

$\Delta([Cr/Fe])$ & 0.03 & 0.00 & 0.01 & 0.03 & 0.26 & 0.26 &  $0.26\pm0.04$ & \\[2pt]

$\Delta([Fe/H])$ & 0.04 & 0.01 & 0.05 & 0.02 & 0.05 & 0.08 & $0.05\pm0.01$ & \\[2pt]
  
$\Delta([Ni/Fe])$ & 0.02 & 0.03 & 0.02 & 0.00 & 0.11 & 0.12 & $0.08\pm0.01$ & \\[2pt] 
\hline
\end{tabular}
}
\end{table*}
Comparing $\sigma_{tot}$ with the observed dispersion $\sigma_{obs}$ in the Table \ref{tab:4} we found a significant difference between $\sigma_{tot}$ and $\sigma_{obs}$ for [O/Fe], [Na/Fe] and [Mg/Fe], as well as we seen in the Figure \ref{fig:7}, suggesting the presence of a intrinsic spread of these elements within cluster. However the comparison with literature will show that while the [Na/Fe] spread is real, the spread in [O/Fe] and [Mg/Fe] are spurious. This is because in the case of O, its lines at 6300 {\AA} is heavily contaminated by the corresponding emission telluric line which is very difficult to remove. Even if we apply sky-correction during the data reduction, this procedure usually does not correct properly strong telluric emission lines, especially in fiber-fed spectrographs. In the case of Mg the lines we used are weak and blended with CN lines. Without knowing C, N, and O abundances it is impossible to get proper abundances from these features. On the other hand, we see a good agreement for the other elements.
\begin{figure}
 \begin{center}
 \includegraphics[width=0.5\textwidth]{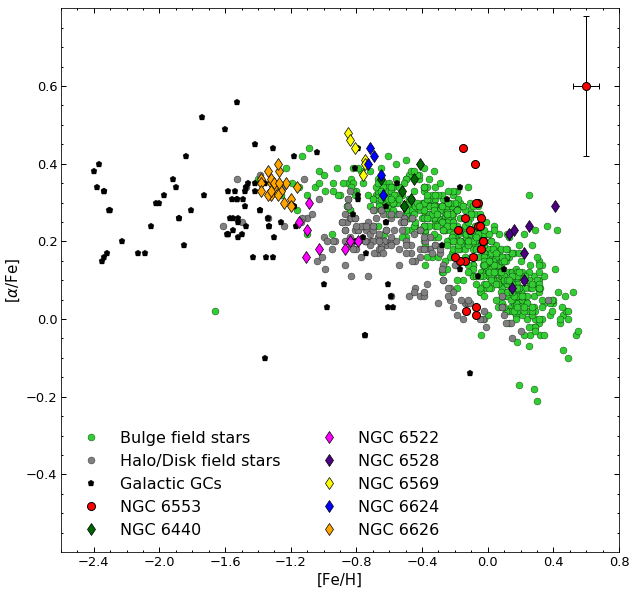}
 \caption{[$\alpha$/Fe] vs [Fe/H]. Filled red circles are our data for NGC 6553. Filled diamonds represent different Bulge GCs, dark green: NGC 6440 \citep{2017MNRAS.469..800M}, magenta: NGC 6522 \citep{2009AA...507..405B}, purple: NGC 6528 \citep{2018AA...620A..96M}, yellow: NGC 6569 \citep{2011MNRAS.414.2690V}, blue: NGC 6624 \citep{2011MNRAS.414.2690V}, orange: NGC 6626 \citep{2017MNRAS.464.2730V}. Filled black pentagon: Galactic GCs \citep{2005AJ....130.2140P}. Filled lime green circles: Bulge field stars \citep{2011AA...530A..54G}. Filled gray circles: Halo and Disk fields stars \citep{2006MNRAS.367.1329R}.} 
 \label{fig:8}
 \end{center}
\end{figure}

\section{Results}

In the following sections we are to present the chemical abundances for Alpha elements, Light elements and Iron peak elements.

\subsection{\texorpdfstring{$\alpha-$}elements}

\begin{figure}
 \begin{center}
 \includegraphics[width=0.49\textwidth, height=0.8\textheight]{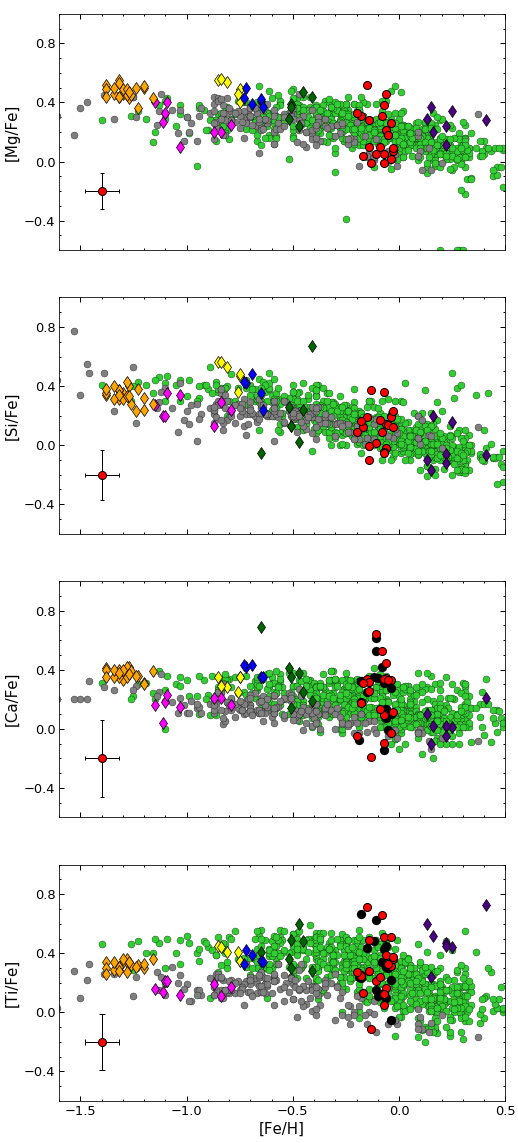}
 \caption{[Mg/Fe], [Si/Fe], [Ca/Fe] and [Ti/Fe] vs [Fe/H]. Filled red and black circles are our data for NGC 6553 derived from spectroscopic  and photometric parameters, respectively. Filled diamonds represent different Bulge GCs, orange: NGC 6626 \citep{2017MNRAS.464.2730V}, magenta: NGC 6522 \citep{2009AA...507..405B}, yellow: NGC 6569 \citep{2011MNRAS.414.2690V}, blue: NGC 6624 \citep{2011MNRAS.414.2690V}, dark green: NGC 6440 \citep{2017AA...605A..12M}, purple: NGC 6528 \citep{2018AA...620A..96M}. Filled lime green circles: Bulge field stars \citep{2011AA...530A..54G}. Filled gray circles: Halo and Disk fields stars \citep{2006MNRAS.367.1329R}.} 
 \label{fig:9}
 \end{center}
\end{figure}

The $\alpha$-elements that we measured in this study are Mg, Si, Ca and Ti (see Table \ref{tab:3} and Figure \ref{fig:9}). Considering Mg, Si, Ca and Ti (O will be considered latter as part of the Na-O anticorrelation) we obtained a mean $\alpha$ abundance of
\begin{center}
 [$\alpha$/Fe] = 0.20$\pm0.02$ dex   
\end{center}
where the error is the error of the mean. This value is in concordance with the value [$\alpha$/Fe] = +0.25 derived by \citet{1999ApJ...523..739C}, where they also use Mg, Si, Ca and Ti to calculate the $\alpha$-elements abundance. Their values, with the exception of Ti which is 0.12 dex lower, are higher than ours, specially in the case of Mg with 0.2 dex of difference. In any case Si and Ca are compatible considering the error. On the other hand, other studies found [$\alpha$/Fe] = +0.13 \citep{2006AA...460..269A} and [$\alpha$/Fe] = +0.11 $\pm0.05$ \citep{2020MNRAS.492.3742M}, lower than our value.

If we compare $\sigma_{tot}$ with $\sigma_{obs}$ (Table \ref{tab:4}) we find no intrinsic spread for the elements Si, Ca and Ti. For [Mg/Fe] see Section \ref{4.2.2}.

The Figure \ref{fig:9} shows the [$\alpha$/Fe] ratio as function of [Fe/H] and due to the apparent large spread present in the abundances of Ca and Ti derived from the spectroscopic parameters, we plotted also the abundances derived from the photometric parameters to evaluate if this bimodality could be due to the atmospheric parameters we used. But, as we see in Figure \ref{fig:9}, the abundances derived from the photometry reflect the same behavior as those derived from the spectra. On the other hand, we can note that all the $\alpha$-elements are overabundant compared to the Sun especially Ti. In any case, we observed that all Bulge GCs show a similar overabundance in the content of their $\alpha$-element.

Figure \ref{fig:8} and Figure \ref{fig:9} show that NGC 6553 is in good agreement with the Bulge field stars and with the other Bulge GCs, indicating that its chemical enrichment was very similar.

\subsection{Light elements}

GCs, unlike the field stars at the same metallicity, show a variation in the content of their light elements abundances (C, N, O, Na, Mg and Al), produced by the Hydrogen burning at high temperatures, which is activated at a temperature of $\thicksim15\times10^{6}$ K, and where the individual abundances of C, N and O are altered, turning C and O into N and by NeNa cycle, which is activated at a temperature of $30\times10^{6}$ K, and where Na is produced at the expense from Ne and finally by the MgAl cycle which is activated at a temperature of $70\times10^{6}$ K, and where, through the capture of protons, Al is produced at the expense of Mg (\citealp{1989ATsir1538...11D,2007AA...470..179P}). 

This variation and the form of N-C, Na-O and Mg-Al (anti)correlations tell us about the MPs phenomenon in GCs.

In NGC 6553 this phenomenon was also observed in the content of  Na, where we noticed an internal spread of this elements (see Table \ref{tab:4}), suggesting that this cluster also presents MPs. In the next section we will discuss this anti-correlation in order to understand the MPs of this cluster.

\defcitealias{2017MNRAS.465...19T}{T17}
\defcitealias{2020MNRAS.492.3742M}{M20}
\begin{figure*}
 \begin{center}
 \includegraphics[width=0.87\textwidth, height=0.47\textheight]{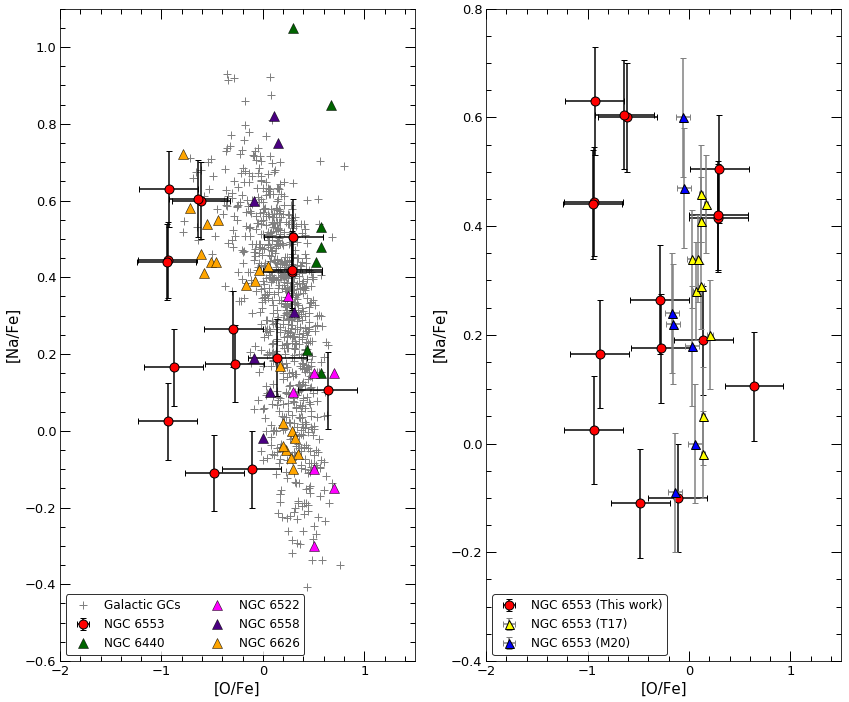}
 \caption{[O/Fe] vs [Na/Fe]. \textbf{Right:} Red filled circles are our data for NGC 6553. Filled triangles represents different Bulge GCs,  dark green: NGC 6440 \citep{2017AA...605A..12M}, magenta: NGC 6522 \citep{2009AA...507..405B}, purple: NGC 6528 \citep{2018AA...620A..96M}, orange: NGC 6626 \citep{2017MNRAS.464.2730V}. Gray crosses: Galactic GCs \citep{2009AA...505..139C}. \textbf{Left:} Red filled circles are our data for NGC 6553. Filled yellow triangle: NGC 6553 \citep{2017MNRAS.465...19T}. Filled blue triangle: NGC 6553 \citep{2020MNRAS.492.3742M}.} 
 \label{fig:10}
 \end{center}
\end{figure*}

\begin{figure}
 \begin{center}
 \includegraphics[width=0.5\textwidth, height=0.41\textheight]{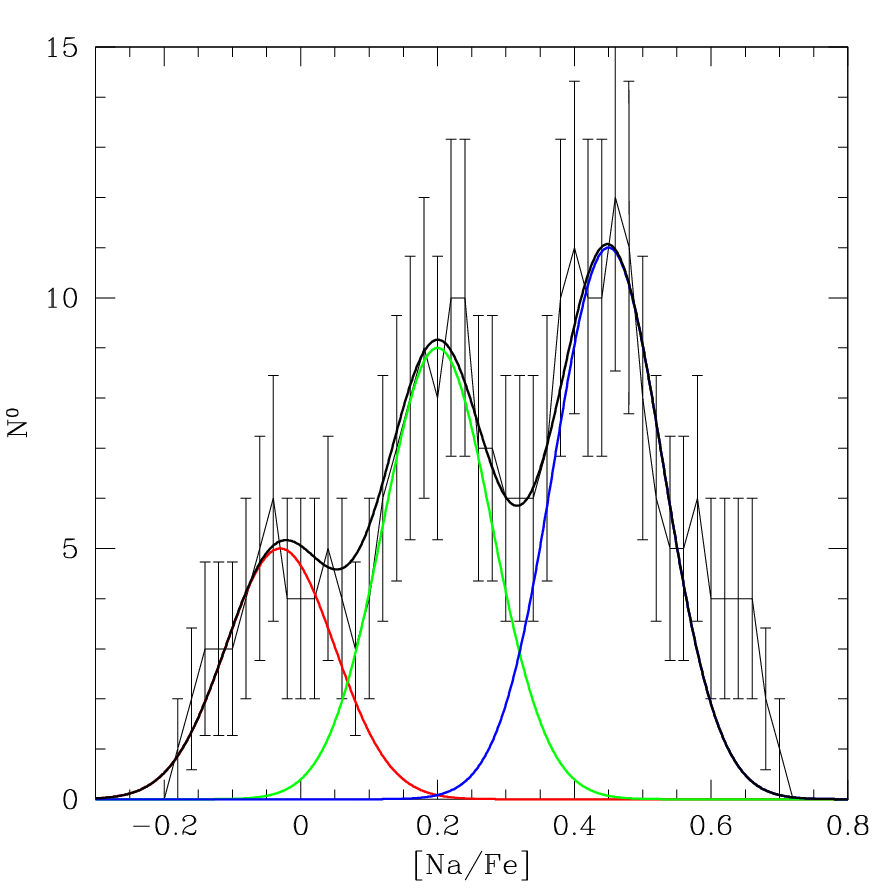}
 \caption{[Na/Fe] distribution for our sample, combined with the samples of \citetalias{2017MNRAS.465...19T} and \citetalias{2020MNRAS.492.3742M}. The continuous black line is the sum of the Gaussians fitting the observational data. The red, green, and blue Gaussians represent each population found in the [Na/Fe] content.} \label{fig:11}
 \end{center}
\end{figure}

\begin{figure}
 \begin{center}
 \includegraphics[width=0.51\textwidth, height=0.40\textheight]{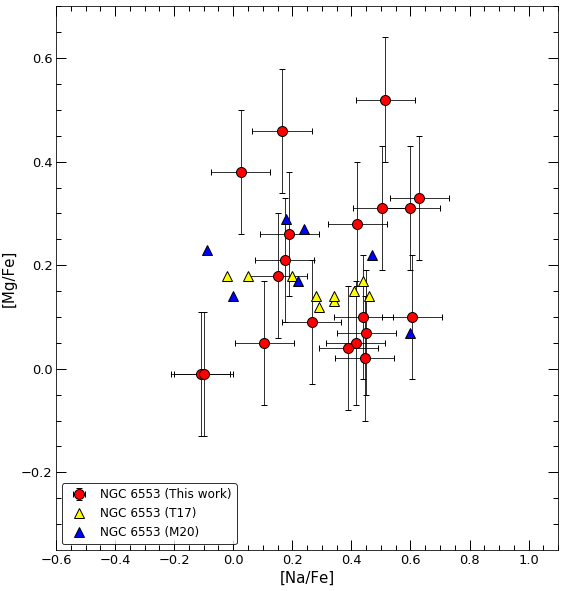}
 \caption{[Na/Fe] vs [Mg/Fe]. Red filled are our data for NGC 6553. Filled yellow triangle: NGC 6553    \citep{2017MNRAS.465...19T}. Filled blue triangle: NGC 6553 \citep{2020MNRAS.492.3742M}.} 
 \label{fig:12}
 \end{center}
\end{figure}

\subsubsection{Na-O anticorrelation}

This is the famous anticorrelation that characterizes almost all GCs, well described by \citet{2009AA...505..117C, 2009AA...505..139C} and \citet{2012AARv..20...50G}. Until now, the only exception to this rule has been Ruprecht 106 \citep{2013ApJ...778..186V}. The Na-O anticorrelation is the most important chemical signature of the presence of MP and the anticorrelation that exists between these two elements within the GC, tells us that there is a first Na-poor and O-rich generation of stars, followed by a second generation Na-rich and O-poor. It is interesting to note that \citet{2007ApJ...671L.125C} found a correlation between the extension of the Na-O anticorrelation observed in RGB stars and the morphology of the HB. They observed that the longer is the O-depleted tail of the Na-O anticorrelation observed in the RGB stars, the higher is the maximum temperature reached by the bluest HB stars in the GC, thus presenting evidence of a link between the morphology of the HB and the presence of star-to-star abundance variations of the elements involved in proton capture.

In NGC 6553 we found a significant spread in Na ($\sigma_{obs}=0.22$) much higher than the total error ($\sigma_{tot}=0.10$). Studies that have searched for MPs in NGC 6553 have been \citet[hereafter \citetalias{2017MNRAS.465...19T}]{2017MNRAS.465...19T} and \citet[hereafter \citetalias{2020MNRAS.492.3742M}]{2020MNRAS.492.3742M}. \citetalias{2017MNRAS.465...19T} used an APOGEE RGB sample, where they obtained an average of [O/Fe] = 0.12 dex and [Na/Fe] = 0.28 dex. Our mean value of O is lower but the Na value is in concordance with its results. In \citetalias{2020MNRAS.492.3742M} they studied 7 RGB stars using UVES spectra and derived a mean of [O/Fe] = -0.07 dex much larger that our value but in the case of Na, they measured [Na/Fe] = 0.23 dex, which is in well agreement with our result considering the errors.

In the Figure \ref{fig:10} we plot [Na/Fe] as function of [O/Fe]. In this figure we see that NGC 6553 present a vertical Na-O anticorrelation with a significant spread in the content of Na and O. While the [Na/Fe] spread we find is confirmed by the other works, this is not true for [O/Fe] since both \citetalias{2017MNRAS.465...19T} and \citetalias{2020MNRAS.492.3742M} do not find any spread in Oxygen. Since at the epoch of our observation the O telluric emission line fell on top of the O 6300 {\AA} line and because in such a case it is very difficult to remove, especially for fiber fed spectrographs, we conclude that the intrinsic O spread we observe is not reliable.

In the left panel of the Figure \ref{fig:10} we compared Na-O anticorrelation of NGC 6553 with other Bulge GCs, as well as with Galactic GCs \citep{2009AA...505..139C}. We note that NGC 6440, NGC 6522 and NGC 6528 show a vertical anticorrelation as well, with a significant spread in Na but not in O. The exception here is NGC 6626 that show the typical Na-O anticorrelation. Even so, this behavior in the Bulge GCs could suggest a different chemical evolution compared with other Galactic GCs.

Analysing the spread of our sample of [Na/Fe], we noticed that there are three groups of stars, which is also seen in the \citetalias{2017MNRAS.465...19T} and \citetalias{2020MNRAS.492.3742M} samples. In order to better distinguish these groups, we plot in the Figure \ref{fig:11} the [Na/Fe] distribution of our sample in combination of with the samples of \citetalias{2017MNRAS.465...19T} and \citetalias{2020MNRAS.492.3742M} and we found that the spread of Na in NGC 6553 follow a distribution where we can see clearly three peaks, represented by the red, green and blue Gaussian respectively. These three groups in the content of [Na/Fe] are the primordial (P), the Intermediate (I) and extreme (E) components defined by \citet{2009AA...505..117C}. The red Gaussian correspond to the P component with a peak at [Na/Fe] = -0.03 dex, the green Gaussian correspond to the I component with a peak at [Na/Fe] = +0.20 dex and the blue Gaussian represents the E component with a peak at [Na/Fe] = +0.45 dex. Then, in order to check our hypothesis of 3 groups of stars with different [Na/Fe] abundances, we performed a $\chi^2$ test assuming firstly a 3 Gaussian fit as shown in Figure \ref{fig:11}, and secondly just a 1 Gaussian fit. The result of this test is $\chi^2 = 5$ for the first hypothesis and $\chi^2 = 23$ for the second. This means that our hypothesis matches much better the Na distribution of the targets.

\subsubsection{Na-Mg anticorrelation} \label{4.2.2}

Sodium and Magnesium are part of the NeNa and MgAl cycles that are activated at different temperatures. The relationship between these two elements tells us about the early internal enrichment that the cluster experienced.

\begin{figure*}
 \begin{center}
 \includegraphics[width=0.97\textwidth, height=0.41\textheight]{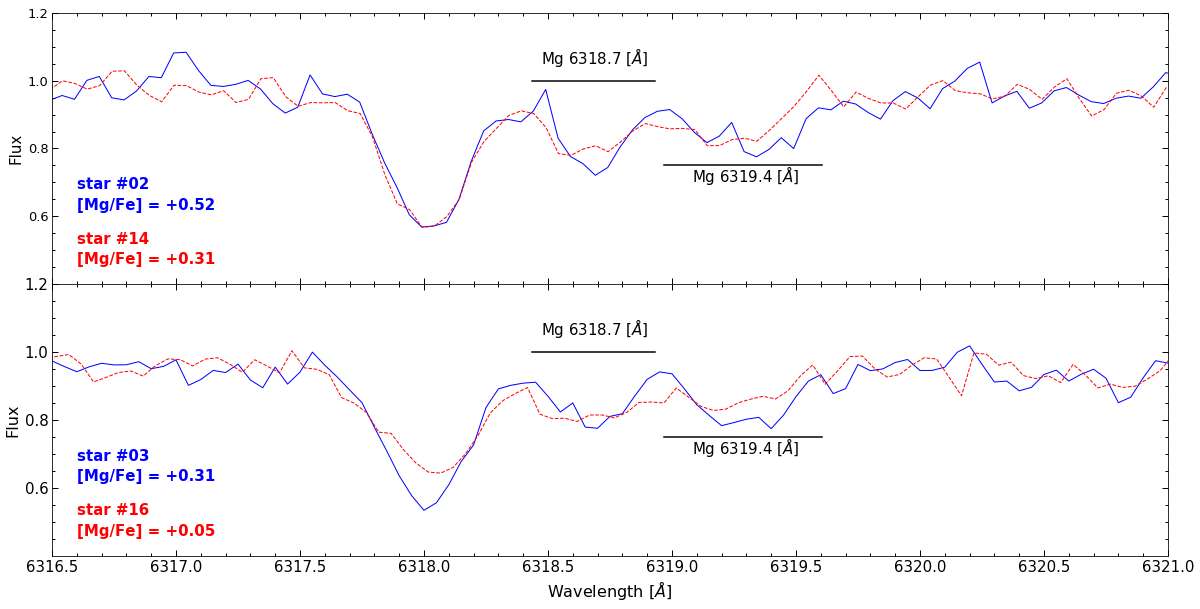}
 \caption{MgI doublet (6318.7 {\AA} and 6319.4 {\AA}) spectral region for two couples of stars with similar atmospheric parameters and very different Mg abundances. The blue full and red dashed line represent the highest and lowest Mg abundances, respectively.} 
 \label{fig:13}
 \end{center}
\end{figure*}

In Figure \ref{fig:12} we have plotted the abundances for [Na/Fe] and [Mg/Fe]. While our data suggest a [Mg/Fe] spread, both \citetalias{2017MNRAS.465...19T} and \citetalias{2020MNRAS.492.3742M} do not support this result. We believe that the spread that we observe in Figure \ref{fig:12} is due to the fact that Mg lines that we measure are weak and for this reason strongly affected by the relative low S/N of the spectra. Figure \ref{fig:13} plots an example of the observed Mg lines for two couples of stars with similar atmospheric parameters and very different Mg abundances. This Figure shows that the spectral variation due to the S/N is the cause of the Mg difference we measured in these stars. On the other hand, we did not find any telluric absorption line falling on the top of out Mg lines, but they are blended with CN lines. 

Additionally, we have decided to assess the impact of CNO variations on Mg abundances. For this, we have separate our sample into first generation (FG) and second generation (SG) stars, using the criteria described in \citet{2009AA...505..117C}, who establish a [Na/Fe] limit for FG stars of $[Na/Fe]_{min} + 4\sigma([Na/Fe])$, where $\sigma([Na/Fe])$ is the star-to-star error on [Na/Fe]. In this way, we obtained 09 FG stars (\#01, \#04, \#05, \#07, \#08, \#13, \#16, \#18 and \#19) and 11 SG stars (\#02, \#03, \#06, \#09, \#10, \#11, \#12, \#14, \#15, \#17 and \#20). The mean values for C and N for FG and SG stars for this cluster were obtained from \citet{2017MNRAS.466.1010S} while the mean value for the O for FG and SG stars, due to the strong telluric contamination of our line at 6300 {\AA}, were obtained from \citetalias{2017MNRAS.465...19T}+\citetalias{2020MNRAS.492.3742M}. We found that the impact of the CNO variation on the Mg lines is negligible. Our conclusion is that the Mg spread we observe is very likely explained by the low S/N of the spectral region were the Mg lines lie.

\subsection{Iron}
 \begin{figure}
 \begin{center}
 \includegraphics[width=0.5\textwidth, height=0.33\textheight]{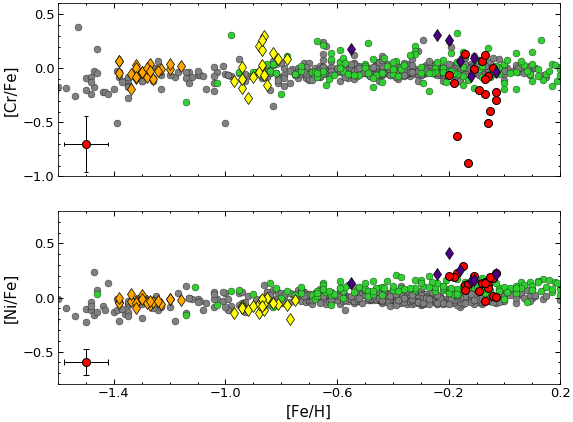}
 \caption{[Cr/Fe], [Ni/Fe] vs [Fe/H]. Red filled circles are our data for NGC 6553. Filled diamonds represents different Bulge GCs, orange: NGC 6626 \citep{2017MNRAS.464.2730V}, yellow: NGC 6569 \citep{2018AJ....155...71J}, purple: NGC 6528 \citep{2018AA...620A..96M}. Filled lime green circles: Bulge field stars \citep{2014AJ....148...67J}. Filled gray circles: Halo and Disk fields stars (\citealp{2003MNRAS.340..304R, 2006MNRAS.367.1329R, 2000AAS..141..491C, 2013ApJ...771...67I}).} 
 \label{fig:14}
 \end{center}
\end{figure}

We found a mean metallicity for NGC 6553 of:
\begin{center}
[Fe/H] = -0.10$\pm0.01$ dex    
\end{center}
We did not find any intrinsic spread in [Fe/H], since the observed scatter is $\sigma_{obs} = 0.05$, that is in good agreement with $\sigma_{tot} = 0.08$.

The measurements of the average metallicity for NGC 6553 in the literature have varied a lot:  \citet{1999AA...341..539B} analysing two stars found an average metallicity of $[Fe/H] = -0.55\pm0.2$ dex. \citet{1999ApJ...523..739C} analysing a sample of five RHB stars found a metallicity of $[Fe/H] = -0.16$ dex. \citet{2006AA...460..269A} analysing a UVES sample spectra of four stars found a metallicity of $[Fe/H] = -0.20$ dex.  \citet{2003AA...411..417M} using IR high- resolution spectra of five stars found a mean metallicity of $[Fe/H] =  -0.20\pm 0.10$ dex. We think that the reason for this discrepancy is the difficulty to identify real members based on radial velocities only. In fact we selected 32 radial velocity members but only 20 of them turned out to be real members when we included proper motions in our analysis. The availability of Gaia data make the study of Bulge clusters much more accurate. Our value is in agreement with the most recently estimates:  \citet{2014AJ....148...67J} using GIRAFFE spectra for twelve stars found an average metallicity of $[Fe/H] = -0.11$ dex with a spread in iron of $\sigma=0.07$, in excellent agreement with our value.  \citetalias{2017MNRAS.465...19T} using a sample of APOGEE spectra found a mean of $[Fe/H] = -0.15\pm0.05$ dex and \citetalias{2020MNRAS.492.3742M} using UVES spectra found a mean of $[Fe/H] = -0.14\pm0.07$ dex and a scatter of $\sigma_{obs}=0.06$ for seven stars. Finally, \citet{2010arXiv1012.3224H} cites a value of $[Fe/H] = -0.18$ dex. Thus, we concluded that our mean metallicity are in concordance with almost all previous work.

\subsection{Iron-Peak Elements}

The iron-peak elements are an important face in the chemical evolution of a GC. They are produced mainly by SNe, which are one of the main polluter of the Interstellar Medium. Here we have measured the abundance of only two iron-peak element: Cr and Ni (see Table \ref{tab:3}), where we found that [Cr/Fe] is underabundant with respect to the Sun but that [Ni/Fe]  is overabundant.

In the Figure \ref{fig:14} we have plotted this iron-peak elements with respect to [Fe/H] comparing them with other Bulge GCs, such as NGC 6626 \citep{2017MNRAS.464.2730V}, NGC 6569 \citep{2018AJ....155...71J} and NGC 6528 \citep{2018AA...620A..96M}. As well as Bulge field stars \citep{2014AJ....148...67J} and Disk and Halo field stars (\citealp{2003MNRAS.340..304R, 2006MNRAS.367.1329R, 2000AAS..141..491C, 2013ApJ...771...67I}). The low [Cr/Fe] abundances observed in some stars is very likely due to the fact that the Cr line is weak and to the relatively low S/N, as in the case of Mg. In the case of [Ni/Fe], we see a good agreement with the Bulge stars.

NGC 6553 is in good agreement with NGC 6528, since they share a similar metallicity ($[Fe/H]_{NGC6528} = -0.14\pm0.03$ dex , \citet{2018AA...620A..96M}) and that follow the spread of the Bulge field star, specially in the case of Ni.

\citetalias{2017MNRAS.465...19T} and \citetalias{2020MNRAS.492.3742M} using high-resolution spectra from APOGEE DR13 and UVES respectively, measured these elements as well. Both \citetalias{2017MNRAS.465...19T} and \citetalias{2020MNRAS.492.3742M} found a mean value of $[Cr/Fe] = 0.00$ dex (with $[Cr/Fe] = 0.00\pm0.03$ in the case of \citetalias{2020MNRAS.492.3742M}), and their value is much larger that the mean value we found ($[Cr/Fe] = -0.20\pm0.06$ dex). In the case of Ni, \citetalias{2017MNRAS.465...19T} found a mean value of $[Ni/Fe] = 0.06$ dex while \citetalias{2020MNRAS.492.3742M} found a mean value of $[Ni/Fe] = 0.26\pm0.03$ dex and in our case, we obtained a mean value of $[Ni/Fe] = 0.13\pm0.02$. All of these mean values differ but they are in concordance if we consider the uncertainties.

\section{The Orbit}

We derived a distance for NGC 6553 of $d = 5.2$ Kpc from the Sun, obtained through the distance modulus:

\begin{equation}
     d = 10^{\frac{(m-M)+5}{5}}    
\end{equation}
Where
\begin{equation}
     (m-M) = (m-M)_{K_{s}} - A_{K_{s}}
\end{equation}
and $A_{K_{s}} = 0.689E(J-K_{s})$ \citep{1989ApJ...345..245C}.
From the isochrone, we obtained $(m-M)_{K_{s}} = 13.8$ and $E(J-K_{s}) = 0.32$. Thus, $(m-M) = 13.58$. This value of distance is in excellent agreement with \citet{2002ASSL..274..107Z}, that found a distance of $d = 5.3$ Kpc.

From the distance, the position and the velocities we calculated the orbit using the GravPot16 code \citep{Fernandez-Trincado}. This is a web code that allow to the user performs a variety of analysis, one of them is the calculate of the orbit, which is based on a Galactic gravitational potential driven by Besan\c con Galaxy Model mass distribution. Thus, to calculate the orbit we assumed the following input parameters for the cluster:
\begin{flushleft}
$R.A._{cluster}=272.323$  $degrees$\\
$DEC_{cluster} = -25.9086$ $degrees$\\
$pm_{R.A.} = 0.2753$ $mas$ $yr^{-1}$\\
$pm_{DEC} = -0.3377$ $mas$ $yr^{-1}$\\
$RV_{helio.} = -3.337$ $km$ $s^{-1}$\\
$d = 5.2$ $kpc$\\
\end{flushleft}

\begin{figure}
 \begin{center}
 \includegraphics[width=0.5\textwidth, height=0.317\textheight]{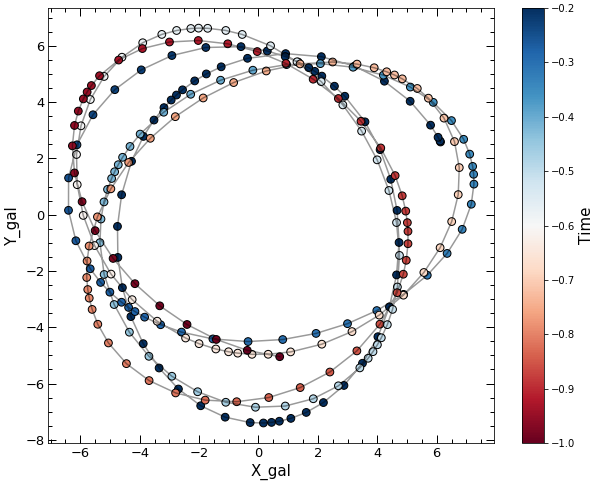}
 \caption{Projection on the X, Y galactic plane of the orbit} 
 \label{fig:15}
 \end{center}
\end{figure}

\begin{figure}
 \begin{center}
 \includegraphics[width=0.5\textwidth, height=0.317\textheight]{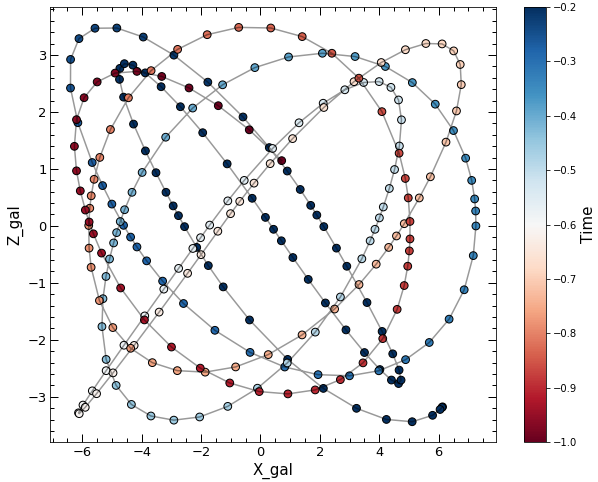}
 \caption{Projection on the X, Z galactic plane of the orbit} 
 \label{fig:16}
 \end{center}
\end{figure}

We integrated the orbit for 1 Gyr. In the Figure \ref{fig:15} we plot the projection of the orbit on the X, Y Galactic Plane. In the Figure \ref{fig:16} we plot the projection of the orbit on the X, Z Galactic Plane. The maximum high above and below the plane is of about 3 Kpcs, while the minimum and maximum distances from the Galactic center are about 4 and 6.5 Kpcs respectively. The orbit does not suggest the cluster as a Bulge member. However this result is not conclusive since an error analysis should be performed before reaching any final conclusion. Also the cluster was formed long before the Bulge had its final shape, so NGC 6553 orbit could reflect this very early stage of the Bulge formation. In any case its chemical content place it as a inner Galaxy object.

\section{Conclusions}

In this work, we have realized a chemical analysis of the Bulge Globular Cluster NGC 6553 to a sample of 116 giant stars both in the horizontal branch as giant branch, using the FLAMES/GIRAFFE spectrograph. In addition, for the first time in NGC 6553, we carried out a proper motion correction using the data release 2 (DR2) granted by the Gaia mission to members stars, obtained by radial velocity, with the aim to determine with great precision the members of the cluster.

Analysing the high-resolution spectra, with atmospheric models and performing the equivalent width method, as well as the spectrum-synthesis (in the cases that required them), we measures the abundance to 9 chemical elements (O, Na, Mg, Si, Ca, Ti, Cr, Fe and Ni) and we perform a detailed study of the errors, comparing how the observed error varied in relation to the total error that we found. Then, we have realized a comparison between NGC 6553  and other nearby Bulge globular clusters, as well as with other galaxy globular clusters and field stars from both Bulge and the Disk and Halo of the Milky Way. In this way, the most important results are detailed in the following:\\

-We found a mean radial velocity of $\langle RV_{H} \rangle  = -3.33\pm1.16$ $Km$ $s^{-1}$, which is in good agreement with the values previously found in the literature.\\

-We found, using radial velocity, 32 stars as member stars, of which 7 were RGB stars and 25 were RHB stars. Then, applying proper motion correction of Gaia, we had 20 stars as final member stars for NGC 6553, all of them being stars of the horizontal branch.\\

-We measure five $\alpha$-elements (O, Mg, Si, Ca and Ti), where all of them are superabundant with respect to the sun, especially in the case of Ti. Considering only the abundances of Mg, Si, Ca and Ti, we find an average of [$\alpha$/Fe] = 0.21$\pm0.02$ dex, which is in good agreement with other data from previous literature, considering the uncertainties. On the other hand, NGC 6553 $\alpha$-elements show good agreement with Bulge field stars.\\

-In the case of the Na-O anticorrelation, we find a vertical  anticorrelation, where there is a spread in Na but the variation in O is minimal. This anticorrelation is very similar to that found in NGC 6528 and other clusters of the Bulge.\\

-We found three peaks in the spread of [Na/Fe] that we identify as the Primordial component of the first generation and the Intermediate and Extreme components of the second generation defined by \citet{2009AA...505..117C}.\\

-We obtained an average iron abundance of [Fe/H] = -0.10$\pm0.01$ dex, which is in good agreement with other studies carried out previously. On the other hand, we did not find a significant spread in the iron content.\\

-In this paper we only measure 2 iron-peak elements, where we find that they are in agreement with the Bulge field stars and that show a very good agreement with NGC 6528 (especially Ni).\\

-In the case of the orbit we suggest that it could reflect the fact that NGC 6553 was formed long before the Bulge had its final shape.\\

\section*{Acknowledgements}

We gratefully acknowledge use of data from ESO program ID 093.D-0286 taken with the VLT-UT2 telescope. We gratefully  the support provided by Direcci\'on de Postgrado UdeC and Departamento de Astronom\'ia Udec. SV gratefully acknowledges the support provided by Fondecyt reg. n. 1170518. CM thanks the support provided by CONICYT through Beca Postdoctorado en el Extranjero convocatoria 2018 folio 74190045. CM also acknowledges support from the Chilean Centro de Excelencia en Astrof\'isica y Tecnolog\'ias Afines (CATA) BASAL grant AFB-170002 and FONDECYT No. 1181797.

\addcontentsline{toc}{section}{Acknowledgements}

\section*{DATA AVAILABILITY}

The spectroscopic data in this article are available from the ESO Science Archive Facility (\url{http://archive.eso.org/wdb/wdb/adp/phase3\_main/form}) with program ID 093.D-0286. The VVV-PSF photometry were provided by \citet{2017MNRAS.464.1874C} and they will are available upon request to the authors with permission of \citet{2017MNRAS.464.1874C}. Lastly, Gaia DR2 is available from the Gaia Archive (\url{https://gea.esac.esa.int/archive/}). 

\addcontentsline{toc}{section}{Data availability}










\bsp	
\label{lastpage}
\end{document}